\documentclass[11pt]{article}

\usepackage{amsmath, amsthm, amssymb, natbib, color, graphicx, hyperref, amsfonts, sectsty, caption, subcaption, tikz, bbm, algorithm, url, capt-of, algpseudocode, lipsum, verbatim, mathtools, tabulary, booktabs, courier, lmodern, longtable, rotating, authblk, multicol}
\usepackage[T1]{fontenc}
\usetikzlibrary{shapes,snakes}

\sectionfont{\rmfamily\mdseries\scshape\normalsize}
\subsectionfont{\rmfamily\mdseries\normalsize}
\subsubsectionfont{\rmfamily\mdseries\normalsize}
\newcommand{\mc}{\mathcal}
\newcommand{\bea}{\begin{eqnarray}}
\newcommand{\eea}{\end{eqnarray}}
\newcommand{\beas}{\begin{eqnarray*}}
\newcommand{\eeas}{\end{eqnarray*}}

\newcommand{\jacc}{\mbox{\scriptsize Jaccard}}
\newcommand{\adad}{\mbox{\scriptsize AA}}
\newcommand{\cooc}{\mbox{\scriptsize cooc}}
\newcommand{\train}{\mbox{\scriptsize train}}

\newcommand{\pred}{\mbox{\scriptsize pred}}

\theoremstyle{definition}

\theoremstyle{definition}

\theoremstyle{definition}

\theoremstyle{definition}

\theoremstyle{definition}

\theoremstyle{remark}

\theoremstyle{definition}

\theoremstyle{definition}

\title{Link prediction for interdisciplinary collaboration via co-authorship network}
\author{Haeran Cho and Yi Yu\footnote{Corresponding Author. Email: y.yu@bristol.ac.uk.  Phone: +44 (0)117 92 87986}}
\affil{School of Mathematics, University of Bristol}

\begin{document}
\maketitle

\abstract
We analyse the Publication and Research (PURE) data set of University of Bristol collected between $2008$ and $2013$. 
Using the existing co-authorship network and academic information thereof, 
we propose a new link prediction methodology, with the specific aim of identifying 
potential interdisciplinary collaboration in a university-wide collaboration network.   

{\bf keywords}: Co-authorship network, interdisciplinary collaboration, bipartite network, link prediction.

\section{Introduction}

Interdisciplinarity has come to be celebrated in recent years 
with many arguments made in support of interdisciplinary research.  
\citet{rylance2015} noted that
\begin{itemize}
\item complex modern problems, such as climate change and resource security, require
many types of expertise across multiple disciplines; 
\item scientific discoveries are more likely to be made on the boundaries between fields,
with the influence of big data science on many disciplines as an example; and
\item encounters with others fields benefit single disciplines and broaden their horizons.
\end{itemize}

In 2015, UK higher education funding bodies and Medical Research Council 
commissioned a quantitative review of interdisciplinary research \citep{elsevier2015}, 
as part of the effort to assess the quality of research produced by UK higher education institutions 
and design the UK's future research policy and funding allocations.  
Around the same time, Nature published a special issue \citep{nature2015}, 
reflecting the increasing trend of interdisciplinarity.  
One such example is observed in publication data,
where more than one-third of the references in scientific papers point to other disciplines; 
also, an increasing number of research centres and institutes established globally, bringing together members of different fields,
in order to tackle scientific and societal questions that go beyond the boundary of a single discipline \citep{ledford2015}.

As a way of promoting interdisciplinary research, \cite{brown2015} suggested 
`the institutions to identify research strengths that show potential 
for interdisciplinary collaboration and incentivise it through seed grants'.
Faced with the problem of utilising limited resources, 
decision makers in academic organisations may focus on
promoting existing collaborations between different disciplines.
However, it could also be of interest to identify the disciplines that have not yet collaborated to this date
but have the potential to develop and benefit from collaborative research given the nurturing environment.

Thus motivated, the current paper has a twofold goal: 
from the perspective of methodological development, we introduce new methods for predicting edges in a network; 
from the policy making perspective, we provide decision makers a systematic way of 
introducing or evaluating calls for interdisciplinary research,
based on the potential for interdisciplinary collaboration detected from the existing co-authorship network.
In doing so, we analyse the University of Bristol's research output data set, which contains 
the co-authorship network among the academic staff and 
information on their academic membership, including the (main) disciplines where their research lies in.


Link prediction is a fundamental problem in network statistics.  
Besides the applications to co-authorship networks, 
link prediction problems are of increasing interests for friendship recommendation in social networks
\citep[e.g.][]{liben2007}, 
exploring collaboration in academic contexts \citep[e.g.,][]{KuzminEtal2016, WangSukthankar2013},
discovering unobserved relationships in food webs \citep[e.g.,][]{WangEtal2014}, 
understanding the protein-protein interactions \citep[e.g.,][]{MartinezEtal2014} 
and gene regulatory networks \citep[e.g.,][]{TurkiWang2015}, to name but a few.  
Due to the popularity of link prediction in a wide range of applications, 
many efforts have been made in developing statistical methods for link prediction problems.  
\cite{liben2007}, \cite{LvZhou2011} and \cite{MartinezEtal2016}, among others, 
are some recent survey papers on this topic.  
The methods developed can be roughly categorised into model-free and model-based methods.  

Among the model-free methods, some are based on information from neighbours 
\citep[e.g.,][]{liben2007, adamic2003, ZhouEtal2009} to form similarity measures and predict linkage; 
some are based on geodesic path information \citep[e.g.,][]{Katz1953, LeichtEtal2006}; 
some use the spectral properties of adjacency matrices \citep[e.g.][]{FoussEtal2007}.
Among the model-based methods, some exploit random walks on the graphs to predict future linkage 
\citep[e.g.,][]{PageEtal1999, JehWidom2002, LiuLv2010};
some predict links based on probabilistic models \citep[e.g.,][]{Geyer1992};
some estimate the network structure via maximum likelihood estimation \citep[e.g.,][]{GuimeraSales2009}; 
others utilise the community detection methods \citep[e.g.,][]{ClausetEtal2008}.

The link prediction problem in this paper shares similarity with the above mentioned ones.
However, we also note on the fundamental difference, that
we collect the data at the level of individual researchers for the large-size network thereof,
but the conclusion we seek is for the small-size network with nodes representing the individuals' academic disciplines, which are given in the data set.
Nodes of the small-size network are different from communities: 
memberships to the communities are typically unknown and 
the detection of community structure is often itself of separate interest,
whereas academic affiliations, which we use as a proxy for academic disciplines, 
are easily accessible and treated as known in our study.

The rest of the paper is organised as follows.
Section~\ref{sec:data} provides a detailed description of the publication and research data set collected at the University of Bristol,
as well as the networks arising from the data.
In Section~\ref{sec:link-pred}, we propose a link prediction algorithm,
compare its performance in combination with varying similiarity measures for
predicting the potential interdisciplinary research links
via thorough study of the co-authorship network, 
and demonstrate the good performance of our proposed method. 
Section~\ref{sec:discussion} concludes the paper.
Appendix provides additional information about the data set.

\section{Data description and experiment setup}
\label{sec:data}

\subsection{Data set}

Publication and Research (PURE) is an online system provided by a Danish company Atira.  
It collects, organises and integrates data about research activity and performance.  
Adopting the PURE data set of research outputs collected between 2008 and 2013 
from the University of Bristol (simply referred to as the `University'),
we focus on journal outputs made by academic staff.  
Each of research outputs and members of academic staff has a unique ID.  
The data set also includes the following information:
\begin{itemize}
\item Outputs' titles and publication dates;
\item Authors' publication names, job titles, affiliations within the University;
\item University organisation structures: 
there are 6 Faculties and each Faculty has a few Schools 
and/or Centres (see Tables~\ref{tab:hier} and \ref{tab:org.key} in Appendix).  
We will refer to the Schools and Centres as the School-level organisations, or simply Schools, in the rest of the paper.
\end{itemize}
Journal information is not provided in the data set, but we obtained this information 
using {\tt rcrossref} \citep{ChamberlainEtal2014}.  

\begin{figure}[h]
\centering
\includegraphics[width = \textwidth]{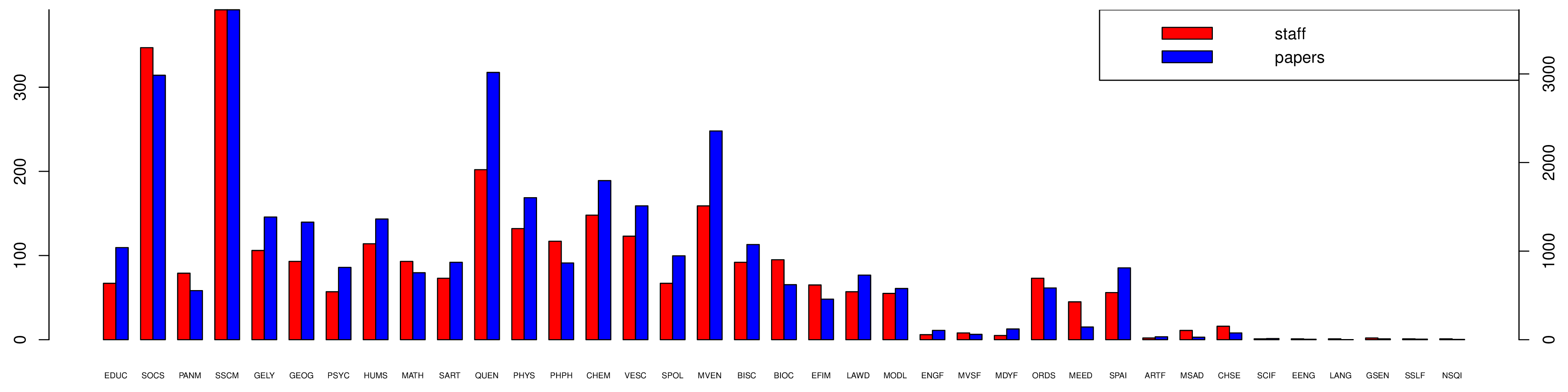}
\caption{Barplot of the number of staff (magnitudes given in the left $y$-axis) and publications (right $y$-axis) from the academic organisations
listed in Table \ref{tab:hier}.}
\label{fig:unbalance}
\end{figure}

In summary, we have
\begin{itemize}
\item 2926 staff, 20 of which have multiple Faculty affiliations, and 	36 of which have multiple School-level affiliations;
\item 20740 outputs, including 3002 outputs in Year 2008, 3084 in 2009, 3371 in 2010, 3619 in 2011, 3797 in 2012, and 3867 in 2013.
\end{itemize}
See Figure~\ref{fig:unbalance} for the breakdown of the academic staff and their publications with respect to the Schools.

\begin{table*}
\begin{center}
\begin{tabular}{p{0.1cm}p{0.1cm}p{0.1cm}p{0.1cm}p{0.1cm}|p{0.1cm}p{0.1cm}p{0.1cm}|p{0.1cm}p{0.1cm}p{0.1cm}|p{0.1cm}p{0.1cm}p{0.1cm}|p{0.1cm}p{0.1cm}p{0.1cm}p{0.1cm}|p{0.1cm}p{0.1cm}p{0.1cm}}
\hline
\multicolumn{18}{c}{UNIV}\\
\hline
\hline
\multicolumn{5}{c|}{FSCI} & \multicolumn{3}{c|}{FSSL} & \multicolumn{3}{c|}{FMVS}  & \multicolumn{3}{c|}{FOAT}  & \multicolumn{4}{c|}{FMDY} & \multicolumn{3}{c}{FENG}\\
\hline
\rotatebox{270}{GELY} & \rotatebox{270}{GEOG} & \rotatebox{270}{PSYC} & \rotatebox{270}{MATH} & \rotatebox{270}{PHYS} & \rotatebox{270}{EDUC} & \rotatebox{270}{LAWD} & \rotatebox{270}{SPAI} &   \rotatebox{270}{PHPH} & \rotatebox{270}{BIOC} & \rotatebox{270}{PANM} & \rotatebox{270}{MODL} & \rotatebox{270}{HUMS} & \rotatebox{270}{SART} & \rotatebox{270}{VESC} & \rotatebox{270}{SOCS} & \rotatebox{270}{ORDS} & \rotatebox{270}{SSCM}  & \rotatebox{270}{QUEN} & \rotatebox{270}{MVEN} & \rotatebox{270}{EENG} \\
\rotatebox{270}{CHEM} & \rotatebox{270}{BISC} & \rotatebox{270}{NSQI} & \rotatebox{270}{SCIF} & & \rotatebox{270}{EFIM} & \rotatebox{270}{SPOL} & \rotatebox{270}{SSLF} & \rotatebox{270}{MSAD} &\rotatebox{270}{MVSF} &  & \rotatebox{270}{LANG} & \rotatebox{270}{ARTF} & & \rotatebox{270}{MEED} & \rotatebox{270}{CHSE} & \rotatebox{270}{MDYF} & & \rotatebox{270}{GSEN} & \rotatebox{270}{ENGF}\\
\hline
\end{tabular}
\caption{Organisation hierarchy structure within the University, full names of which can be found in Table~\ref{tab:org.key} in Appendix.  \label{tab:hier}}
\end{center}
\end{table*}

Note that this data set only includes all the authors within the University, 
i.e., if a paper has authors outside the University, (disciplines of) these authors are not reflected 
in the data set nor the analysis conducted in this paper. 
Also, we omit from our analysis any contribution to books and anthologies, conference proceedings and software.
In Summer 2017, the University has re-named the Schools in the Faculty of Engineering 
and Faculty of Health Sciences, and merged SOCS and SSCM 
as Bristol Medical School (see Table~\ref{tab:org.key}).  
In this paper, we keep the structure and names used for the data period.  

\subsection{Experiment setup and notation}

In order to investigate the prediction performance of the proposed methods, 
we split the whole data set into training and test sets, which contain the research outputs published 
in Years 2008--2010 and Years 2011--2013, respectively.  

Denote by $\mathcal{I}$ and $\mathcal{O}$ the collections of all the staff (researchers) and 
all the School-level organisations appearing in Years 2008--2013, respectively.  
Also, let $\mc J$ denote the collection of all the journals in which the researchers in $\mc I$ have published during the same period.
Three types of networks arise from the PURE data set.
\begin{itemize}
\item Co-authorship network: the nodes are individual researchers ($\mc I$), 
and the edges connecting pairs of researchers indicate that they have joint publications.
\item Researcher-journal network: in this bipartite network, the nodes are researchers ($\mc I$) and journals ($\mc J$),
and there is an edge connecting a researcher and a journal if the researcher has published in the journal. 
\item School network: the nodes are School-level organisations ($\mc O$), 
and the edges connecting pairs of organisations indicate that they have collaboration in ways which are to be specified; we wish to predict links in this network. 
\end{itemize}

The co-authorship adjacency matrices for the training and test sets are denoted by $A^{\mathrm{train}}, A^{\mathrm{test}} \in \mathbb{N}^{|\mathcal{I}| \times |\mathcal{I}|}$, both of which are based on the same cohort of researchers.  To be specific, for $i, j \in \mathcal{I}$,
	\[
	A^{\mathrm{train(test)}}_{ij} = \begin{cases}
		\mbox{number of joint publications between } i \mbox{ and } j \mbox{ in training (test) set}, & i 
		\neq j; \\
		0 & i = j.
	\end{cases}
	\]
Similarly, we define the incidence matrices corresponding to the research-journal bipartite networks for the training and test sets
$I^{\mathrm{train}}, I^{\mathrm{test}} \in \mathbb{N}^{|\mathcal{I}| \times |\mathcal{J}|}$, as
\beas
I^{\mathrm{train(test)}}_{ij} = \mbox{ number of publications in journal $j$ by researcher $i$ in training (test) set},
\eeas
for $i \in \mathcal{I}$ and $j \in \mathcal{J}$. 

For a researcher $i \in \mathcal{I}$, let $\mathcal{S}(i)$ be the School-level affiliation of researcher $i$.  
At the School-level, we create collections of edges (collaboration) $E^{\mathrm{train}}$ and $E^{\mathrm{test}}$ for the training and test sets, respectively, with
\[
(k, l) \in E^\text{train(test)} \iff \exists i, j \in \mathcal{I} \text{ such that } s(i) = k, \, s(j) = l \text{ and } 
A_{ij}^\text{train(test)} > 0,
\]
i.e., we suppose that there is an edge connecting a pair of organisations if they have joint publications in the corresponding data sets.  
Note that since $A^{\text{train(test)}}$ are symmetric, the edges in $E^{\text{train(test)}}$ are undirected ones.  
	
Then, $E^{\mathrm{new}} = E^{\mathrm{train}} \setminus E^{\mathrm{test}}$
denotes the collection of new School-level collaborative links appearing in the test set only.	 
In this data set, there are 260 pairs of Schools which have no collaborations in the training set, 
and $|E^{\mathrm{new}}| = m_{\mathrm{new}} = 37$ new pairs of Schools which have developed collaborations in the test set.
Our aim is to predict as many edges in $E^{\mathrm{new}}$ as possible using the training set, 
without incurring too many false positives.  We would like to point out that false positives can also be interpreted as potential collaboration which has not be materialised in the whole data set.

\section{Link prediction}
\label{sec:link-pred}

\subsection{Methodology}
\label{sec:method}

We formulate the problem of predicting potential interdisciplinary collaboration in the University
as School network link prediction problem, by regarding the academic affiliations as a proxy for disciplines.
We may approach the problem
\begin{itemize}
\item[(i)] by observing the potential for future collaboration among the individuals 
and then aggregating the scores according to their affiliations for link prediction in the School network, or

\item[(ii)] by forming the School network based on the existing co-authorship network 
(namely, $(\mc O, E^{\train})$) and predicting the links thereof.
\end{itemize}
Noting that interdisciplinary research is often led by individuals of strong collaborative potential,
we adopt the approach in (i) and propose the following algorithm.

\begin{description}
\item [Link prediction algorithm]
\item [Step 1] Obtain the similarity scores for the pairs of individuals as $\{w^0_{ij}; \, i, j \in \mathcal{I}\}$ using the training data.
\item [Step 2] Assign weights $w_{kl}$ to the edges in the School network by aggregating $w^0_{ij}$
for $i$ with $\mathcal{S}(i) = k$ and $j$ with $\mathcal{S}(j) = l$.
\item [Step 3] Select the set of predicted edges as 
\beas
E^{\mathrm{pred}} = \{(k, l):\, w_{kl} > \pi \mbox{ and } (k, l) \notin E^{\mathrm{train}}\},
\eeas
for a given threshold $\pi$.
\end{description}
Note that, although we can compute the edge weights for the pairs of individuals (and hence for the pairs of Schools) 
with existing collaborative links in Steps 1--2, they are excluded in the prediction performed in Step 3.

We propose two different methods for assigning the similarity scores $w^0_{ij}$ to the pairs of individual researchers in Step 1,
and aggregating them into the School network edge weights $w_{kl}$ in Step 2.
We first compute $w^0_{ij}$ using the co-authorship network only (Section \ref{sec-nei}),
and explore ways of further integrating the additional layer of information by
adopting the bipartite network between the individuals and journals (Section \ref{sec-similarity}).

\subsubsection{Similarity scores based on the co-authorship network}
\label{sec-nei}

As noted in \cite{ClausetEtal2008}, neighbour- or path-based methods 
have been known to work well in link prediction for strongly assortative networks 
such as collaboration and citation networks.  If researchers A and B have both collaborated with researcher C in the past, 
it is reasonable to expect the collaboration between A and B if they have not done so yet.  
In the same spirit, one can also predict linkage based on other functions of neighbourhood.  

Motivated by this observation, we propose different methods for calculating the similarity scores in Step 1.
In all cases, $w_{ij}^0 = 0$ if and only if $(i, j)$ does not have a length-2 
geodesic path based on $A^{\mathrm{train}}$.
\begin{itemize}
\item [(a)] Length-2 geodesic path.  Set $w^0_{ij} = 1$ if there is a length-2 geodesic path connecting $i$ and $j$ based on $A^{\mathrm{train}}$.  
\item [(b)] Number of common direct neighbours.  Let $w^0_{ij}$ be the number of distinct length-2 geodesic paths linking $i$ and $j$ based on $A^{\mathrm{train}}$, i.e.,
	\[
	w^0_{ij} = \mid \mathcal{N}^{\mathrm{train}}(i) \cap \mathcal{N}^{\mathrm{train}}(j)\mid, 
	\]
	where $\mathcal{N}^{\mathrm{train}}(i) = \{k: A^{\mathrm{train}}_{ik} > 0\}$.
\item [(c)]	Number of common order-2 neighbourhood.  Let $w^0_{ij}$ be the number of common order-2 neighbours of $i$ and $j$; 
in other words, 
	\begin{align*}
	 w^0_{ij} = & \mid \bigl(\mathcal{N}^{\mathrm{train}}(i) \cup \{k: \, k\in \mathcal{N}^{\mathrm{train}}(l), \, l \in \mathcal{N}^{\mathrm{train}}(i)\}\bigr) \\
	\cap & \bigl(\mathcal{N}^{\mathrm{train}}(j) \cup \{k: \, k\in \mathcal{N}^{\mathrm{train}}(l), \, l \in \mathcal{N}^{\mathrm{train}}(j)\}\bigr) \mid.
	\end{align*}
\item [(d)] Sum of weights of path edges.  Let $w^0_{ij}$ be the sum of the $A^{\mathrm{train}}$ weights of all the length-2 geodesic paths linking $i$ and $j$, i.e., listing all length-2 geodesic paths connecting $i$ and $j$ as $\{i, k_1, j\}, \{i, k_2, j\}, \cdots, \{i, k_m, j\}$, $m \geq 1$, we set
	\[
	w^0_{ij} = \sum_{s = 1}^m (A^{\mathrm{train}}_{i, k_s} + A^{\mathrm{train}}_{k_s, j}).
	\]
\end{itemize}

All (a)--(d) assign positive weights to the pairs of individuals 
who do not have direct collaboration in the training data set, but have at least one common co-author.  
Compared to (a), the other three scores integrate more information and take into consideration 
the number of common publications or the number of common co-authors; 
however, all (a)--(d) assign non-zero weights to the same set of edges. 
Then, with the thus-chosen edge weights between the researchers, we obtain
the edge weights for the School network in Step 2, as
\beas
w_{kl} = \sum_{i:\, \mathcal{S}(i) = k}\sum_{j:\, \mathcal{S}(j) = l} w^0_{ij} \quad \text{for} \quad k, l \in \mathcal{O},
\eeas
which in turn is used for link prediction in Step 3.
In combination with (a)--(d), 
we propose to select the threshold $\pi$ in Step 3
as the $100(1-p)$th percentile of $\{w_{kl} > 0, k, l \in \mathcal{O}\}$ for a given $p \in [0, 1]$.

\subsubsection{Similarity scores based on the bipartite network}
\label{sec-similarity}

In the research output dataset, we have additional information, 
namely the journals in which the research outputs have been published, 
which can augment the co-authorship network for School network link prediction. 
Our motivation comes from the observation that when researchers from different organisations 
publish their research outputs in the same (or similar) journals but have not collaborated yet to this date, 
it indicates that they have the potential to form interdisciplinary collaboration with each other.  
A similar idea has been adopted in e.g., \cite{KuzminEtal2016} for identifying the potential for 
scientific collaboration among molecular researchers, 
by adding the layer of the paths of molecular interactions to the co-authorship network. 

Recall the incidence matrix for the researcher-journal bipartite network in the training set, $I^{\train}$.
In the bipartite network, we define the neighbours of the researcher $i$ as the journals in which $i$ has published, 
and denote the set of neighbours by $\mc J^{\train}(i) = \{j \in \mathcal{J}: \, I^{\train}_{ij} \ne 0\}$.  
Analogously, for journal $j$, its neighbours are those researchers who have published in the journal, 
and its set of neighbours is denoted by $\mc I^{\train}(j) = \{i \in \mathcal{I}: \, I^{\train}_{ij} \ne 0\}$.

Then, we propose the following scores to be used in Step 1
for measuring the similarity between two researchers $i$ and $i'$.
Where there is no confusion, we omit `train' from the superscripts of $\mc J^{\train}(\cdot)$, $\mc I^{\train}(\cdot)$ and $I^{\train}$.
\begin{description}
\item[Jaccard's coefficient:] The Jaccard coefficient that measures the similarity between finite sets,
is extended to compare the neighbours of two individual researchers as
\beas
\sigma_{\jacc}^1(i, i') = \frac{\vert \mc J(i) \cap \mc J(i') \vert}{\vert \mc J(i) \cup \mc J(i') \vert}.
\eeas
This definition simply counts the number of journals shared by $i$ and $i'$,
and hence gives more weights to a pair of researchers who e.g., 
each published one paper in two common journals, than those who
published multiple papers in a single common journal, given that $\vert \mc J(i) \cup \mc J(i') \vert$ remains the same.
Therefore, we propose a slightly modified definition which takes into account the number of publications:
\beas
\sigma_{\jacc}^2(i, i') = \frac{\sum_{j \in \mc J(i) \cap \mc J(i')} (I_{ij} + I_{i'j})}
{\sum_{j \in \mc J(i) \cup \mc J(i')} (I_{ij} + I_{i'j})}.
\eeas

\item[\cite{adamic2003}:] 
The rarer a journal is (in terms of total publications made in the journal),
two researchers that share the journal may be deemed more similar.
Hence we adopt the similarity measure originally proposed in \cite{adamic2003}
for measuring the similarity between two personal home pages based on the common features,
which refines the simple counting of common features by weighting rarer features more heavily:
\beas
\sigma_{\adad}(i, i') = \sum_{j \in \mc J(i) \cap \mc J(i')} \frac{1}{\log(\sum_{l \in \mc I(j)} I_{lj})}
\eeas

\item[Co-occurrence:] We note the resemblance between the problem of edge prediction in a co-authorship network and that of stochastic language modelling for unseen bigrams (pairs of words that co-occur in a test corpus but not in the training corpus), and adapt the `smoothing' approach of \cite{essen1992}.  We first compute the similarity between journals using $\sigma^k_{\jacc}, \, k = 1, 2$ and augment the similarity score between a pair of researchers by taking into account not only those journals directly shared by the two, but also those which are close to those journals:
\beas
\sigma^k_{\cooc}(i, i') &=& \sum_{j \in \mc J(i)}\sum_{j' \in \mc J(i')} 
\frac{I_{ij}}{\sum_l I_{il}} \cdot \frac{I_{i'j'}}{\sum_l I_{i'l}} \cdot \sigma^k_{\jacc}(j, j'), \quad k = 1, 2.
\eeas
\end{description}
The use of above similarity measures and others have been investigated by \cite{liben2007} for link prediction problems in social networks.
Here, we accommodate the availability of additional information beside the direct co-authorship network, 
and re-define the similarity measures accordingly.

Since the above similarity measures do not account for the path-based information in the co-authorship network,
we propose to aggregate the similarity scores and produce the School network edge weights (Step 2) as
\bea
\label{pred:journal:weight}
w_{kl} = \sum_{i: \, \mc S(i) = k} \sum_{i': \, \mc S(i) = l} \sigma(i, i') \cdot \mathbb{I}(g_{ii'} < d), \qquad k, l \in \mc O,
\eea
for a given $d > 0$,
where $g_{ii'}$ denotes the geodesic distance between researchers $i$ and $i'$ in $A^{\train}$.
As an extra parameter $d$ is introduced in computing $w_{kl}$,
we propose to select the threshold $\pi$ in Step 3 such that only those $(k, l) \notin E^{\train}$,
whose edge weights $w_{kl}$ exceed the median of the 
weights for the collaborative links that already exist in the training set, are selected in $E^{\pred}$.

\subsection{Results}

In Table~\ref{tab:nei-pred}, we perform link prediction following Steps 1--3 of the link prediction algorithm on the PURE data set,
using different combinations of the weights (a)--(d) and the threshold chosen with $p \in \{1, 0.4, 0.3, 0.2\}$ 
as described in Section \ref{sec-nei}, and
similarity scores introduced in Section \ref{sec-nei} together with $d \in \{\mbox{NA}, \infty, 10, 4\}$ for \eqref{pred:journal:weight},
where NA refers to the omission of thresholding on the geodesic distance $g_{ii'}$.
For evaluating the quality of the predicted links,
we report the total number of predicted edges, their prediction accuracy and recall, which are defined as
\begin{align*}
\mbox{prediction accuracy} &= \frac{\mbox{\# of correctly predicted edges}}{\mbox{\# of predicted edges}},
\\
\mbox{recall} &= \frac{\mbox{\# of correctly predicted edges}}{\mbox{\# of all new edges ($m_{\mathrm{new}}$)}},
\end{align*}
following the practice in the link prediction literature (see \cite{liben2007}).
Each method is compared to random guessing, the prediction accuracy of which is defined as 
the expectation of prediction accuracy of randomly picking $m_{\mathrm{new}}$ pairs from all non-collaborated pairs in the training data.

\begin{table}[htbp]
\centering
\caption{Summary of the links predicted with the similarity measures 
and the thresholds chosen with $p \in \{1, 0.4, 0.3, 0.2\}$ as described in Section \ref{sec-nei},
and those described in Section \ref{sec-similarity} with $d \in \{\mbox{NA}, \infty, 10, 4\}$,
in comparison with the links predicted by a modularity-maximising community detection method (comm. detect.)
with varying number of communities $N$.
There are 37 pairs of Schools which have developed new collaborations in the test set,
out of 260 pairs that have no collaborations in the training set.}
\label{tab:nei-pred}
{\scriptsize
\begin{tabular}{c|c|cccc||c|ccccc||c|c}
\hline\hline
&	\multicolumn{5}{c}{Section \ref{sec-nei}} &	 \multicolumn{6}{c}{Section \ref{sec-similarity}} &						
\multicolumn{2}{c}{comm. detect.} 		\\	
&	$p$ &	(a) &	(b) &	(c) &	(d) &	$d$ &	$\sigma^1_{\jacc}$ &	$\sigma^2_{\jacc}$ &	$\sigma_{\adad}$ &	$\sigma^1_{\cooc}$ &	$\sigma^2_{\cooc}$ 	
& $N$ &		\\	\hline
\# of edges &	1 &	80 & 	80 & 	80 & 	80 & 	NA &	43 & 	45 & 	44 & 	33 & 	28 & 	5 &	31	\\	
accuracy &	&	.338 &	.338 &	.338 &	.338 &	&	.488 &	.489 &	.432 &	.606 &	.679 &	&	0.129	\\	
recall &	&	.365 &	.365 &	.365 &	.365 &	&	.284 &	.298 &	.257 &	.270 &	.257 &	&	0.054	\\	\hline
\# of edges &	0.4 &	49 & 	32 & 	32 & 	33 & 	$\infty$ &	20 & 	18 & 	26 & 	26 & 	17 & 	6 &	25	\\	
accuracy &	&	.388 &	.500 &	.469 &	.424 &	&	.650 &	.667 &	.615 &	.769 &	.824 &	&	0.160	\\	
recall &	&	.257 &	.216 &	.203 &	.189 &	&	.176 &	.162 &	.217 &	.270 &	.189 &	&	0.054	\\	\hline
\# of edges &	0.3 &	24 & 	24 & 	24 & 	25 & 	10 &	18 & 	18 & 	23 & 	27 & 	17 & 	7 &	24	\\	
accuracy &	&	.541 &	.583 &	.500 &	.480 &	&	.667 &	.722 &	.652 &	.704 &	.824 &	&	0.166	\\	
recall &	&	.176 &	.189 &	.162 &	.162 &	&	.162 &	.176 &	.203 &	.257 &	.189 &	&	0.050	\\	\hline
\# of edges &	0.2 &	24 & 	16 & 	16 & 	21 & 	4 &	4 & 	4 & 	5 & 	16 & 	5 & 	8 &	21	\\	
accuracy &	&	.541 &	.625 &	.586 &	.523 &	&	.500 &	.750 &	.800 &	.688 &	.800 &	&	0.095	\\	
recall &	&	.176 &	.135 &	.122 &	.149 &	&	.027 &	.041 &	.054 &	.149 &	.054 &	&	0.027	\\	\hline
	\multicolumn{6}{l}{random guess accuracy: $0.142$}  													\\	\hline\hline
\end{tabular}}
\end{table}

\begin{figure}[htbp] 
	\begin{minipage}[b]{0.45\linewidth}
    \centering
    \includegraphics[width=0.825\linewidth]{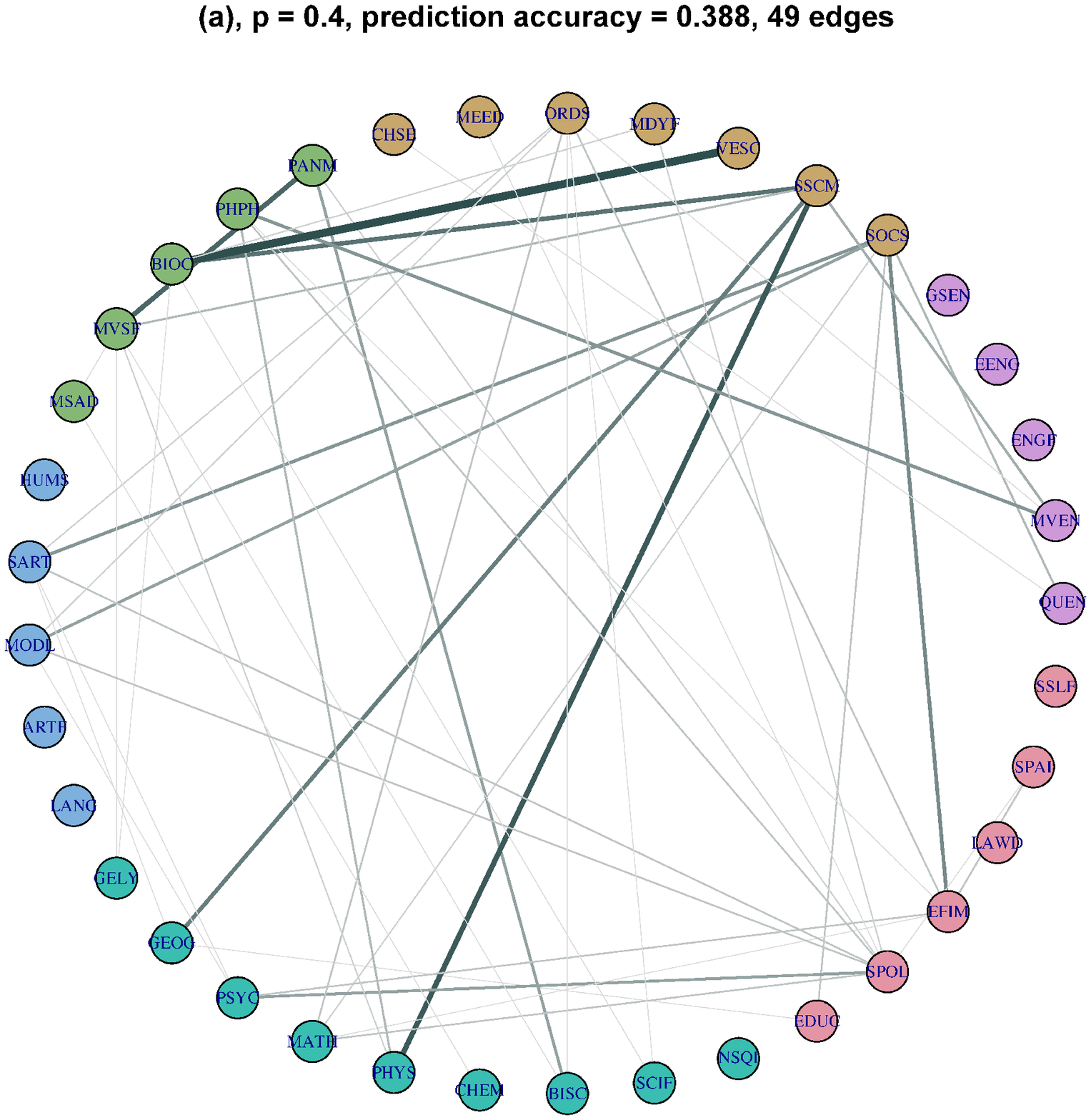} 
    \vspace{4ex}
  \end{minipage}
  \begin{minipage}[b]{0.45\linewidth}
    \centering
    \includegraphics[width=.825\linewidth]{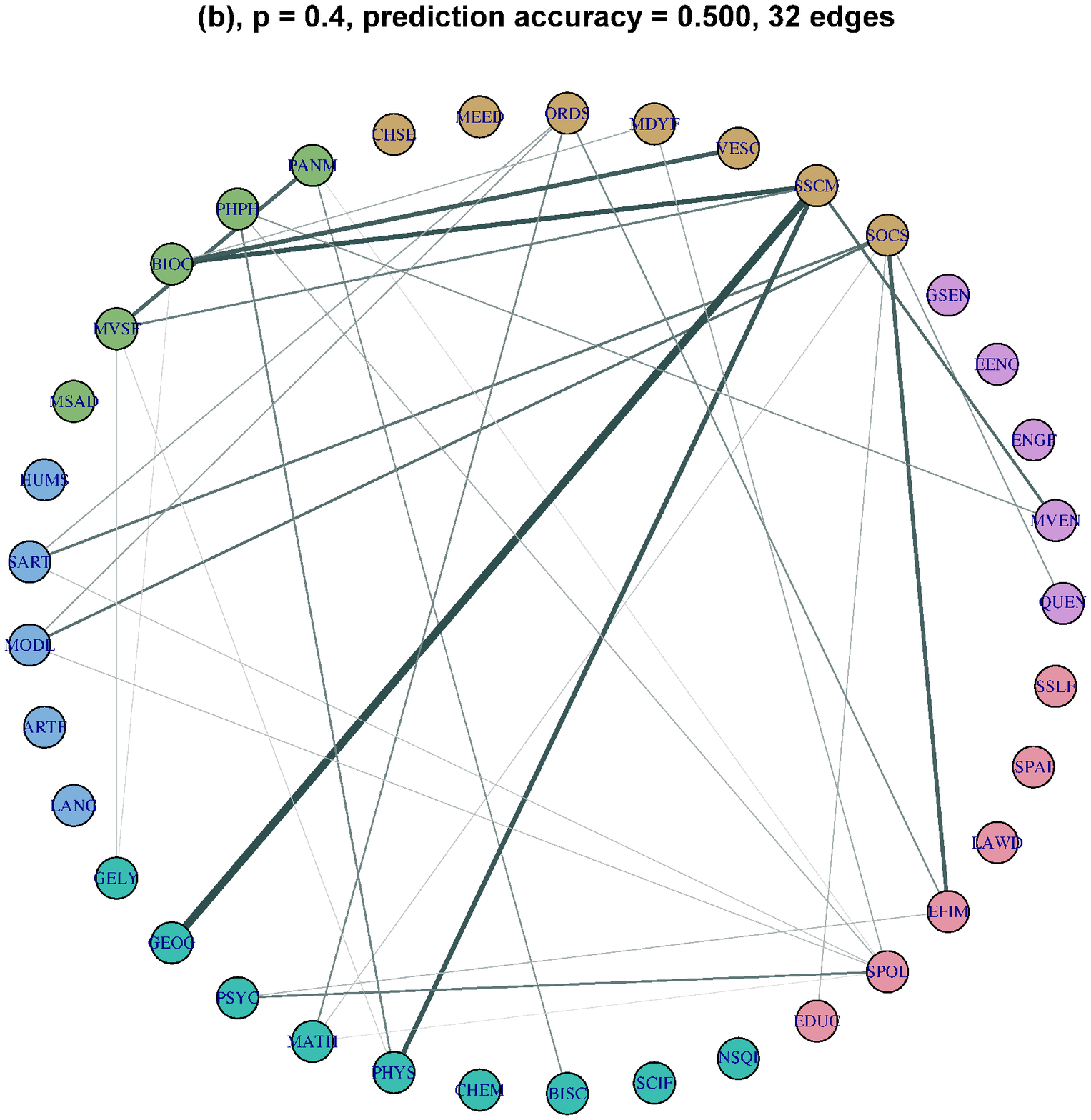} 
   \vspace{4ex}
  \end{minipage}   
  \begin{minipage}[b]{0.45\linewidth}
    \centering
    \includegraphics[width=.825\linewidth]{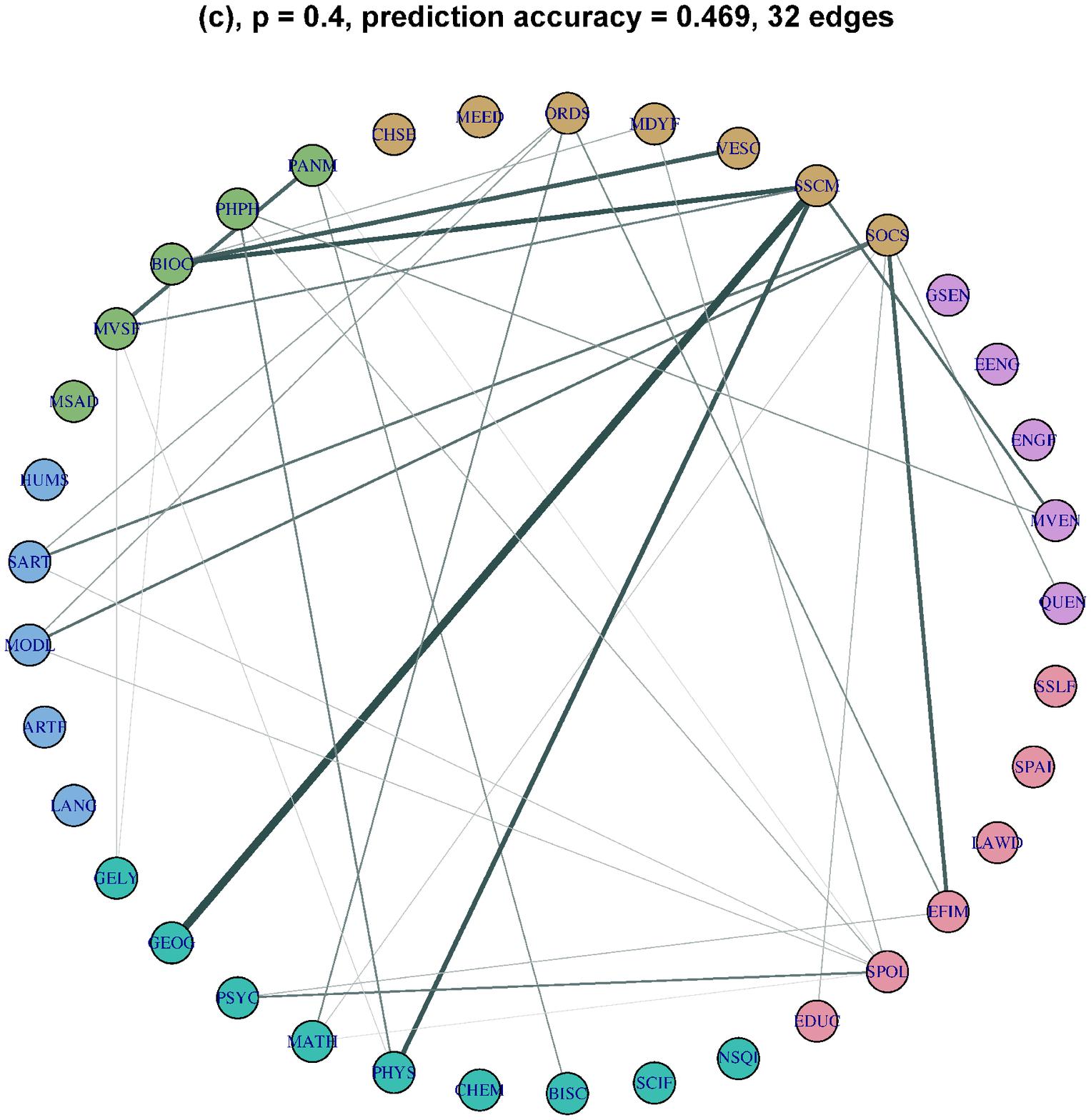} 
  \end{minipage}
    \begin{minipage}[b]{0.45\linewidth}
    \centering
    \includegraphics[width=.825\linewidth]{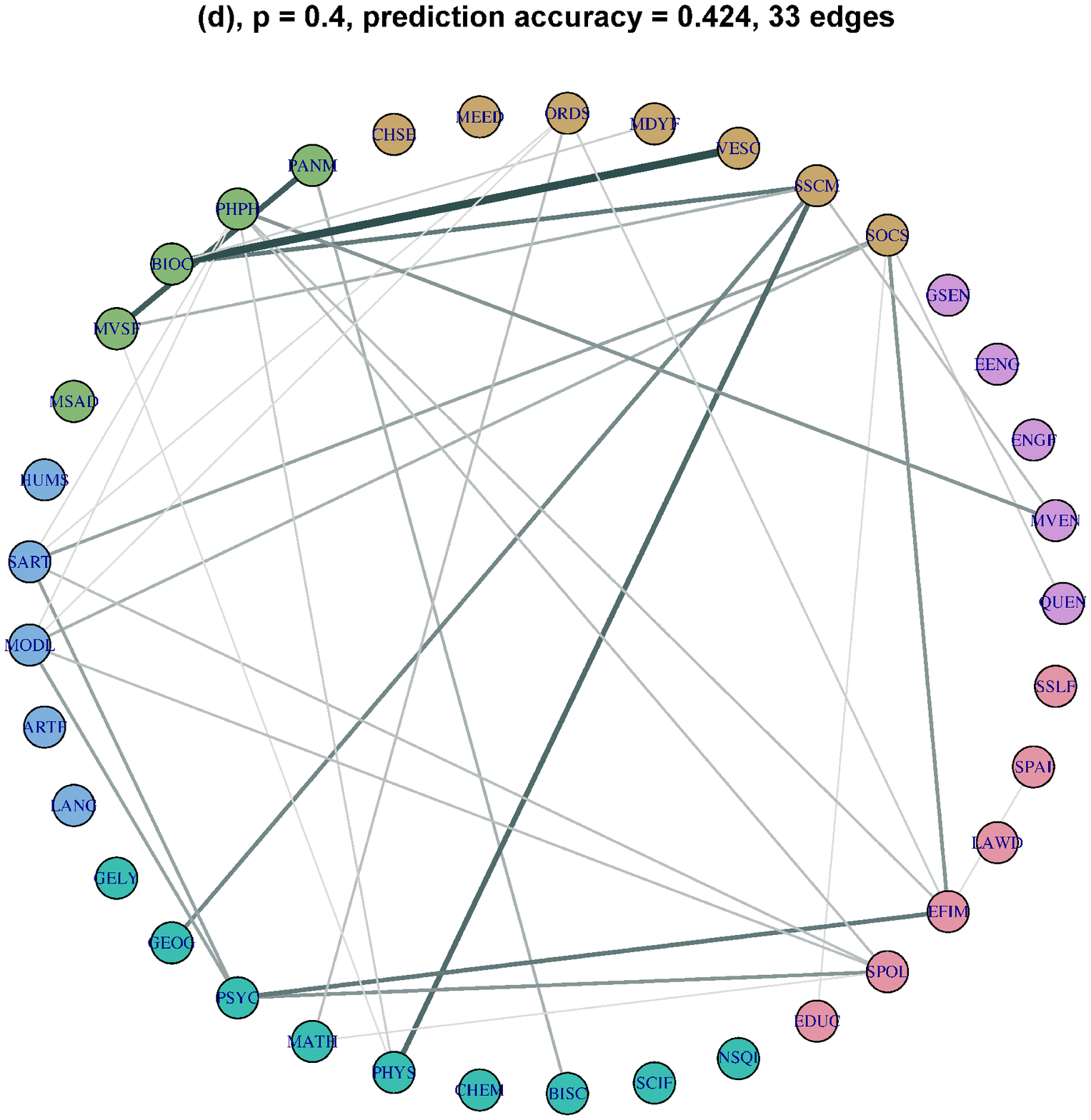} 
  \end{minipage} 
\caption{Edges predicted indicating possible collaboration among School-level organisations using various weights (a)--(d) described in Section \ref{sec-nei} and threshold $p = 0.4$.  Each node represents a School and each Faculty has a unique colour. Each plot reports the prediction accuracy and the number of total edges returned. The edge width is proportional to the edge weights $w_{kl}$ in Step 1.}
\label{fig:nei:pred}  
\end{figure}

\begin{figure}[htbp] 
	\begin{minipage}[b]{0.5\linewidth}
    \centering
    \includegraphics[width=.825\linewidth]{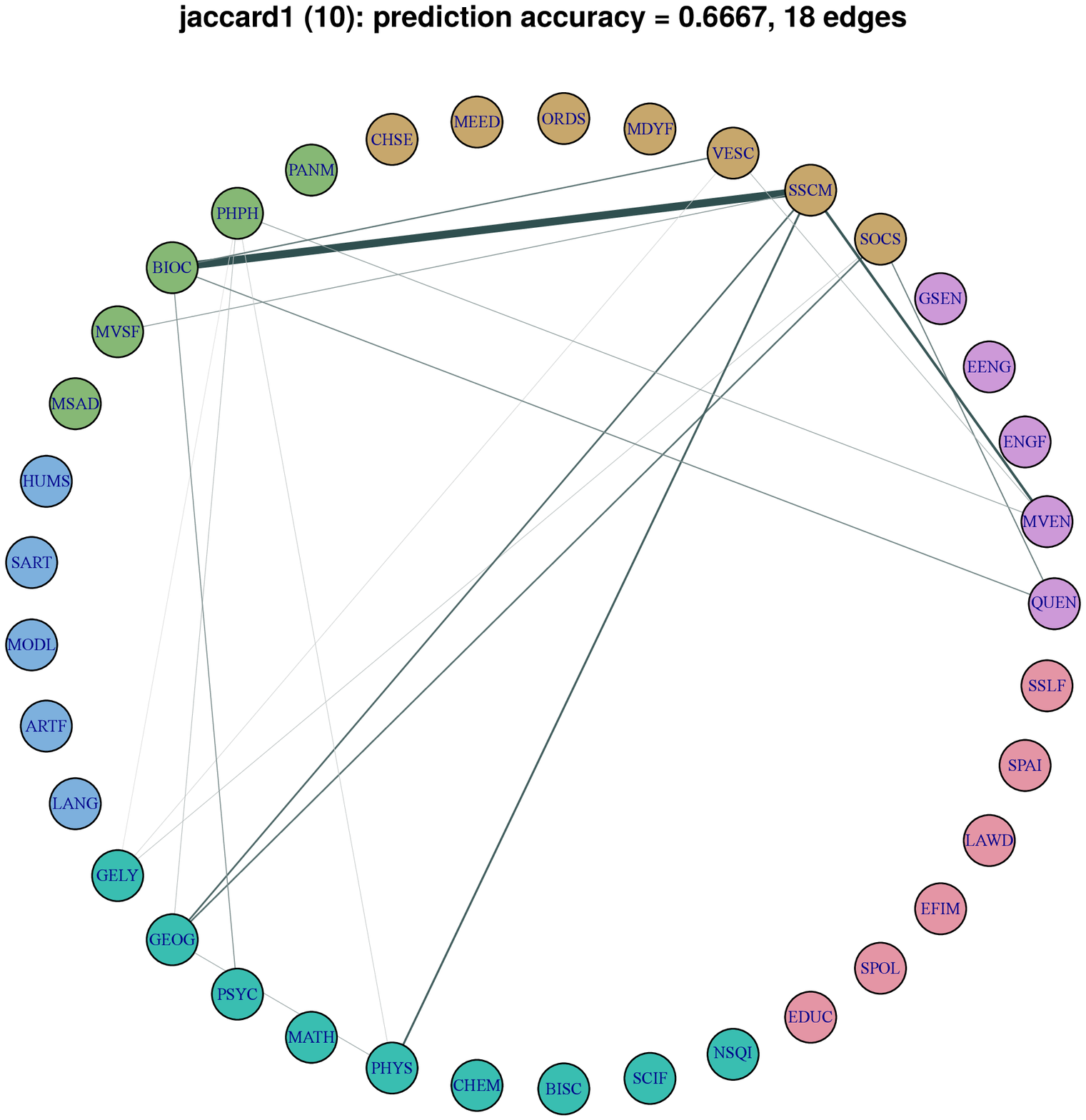} 
    \vspace{4ex}
  \end{minipage}
  \begin{minipage}[b]{0.5\linewidth}
    \centering
    \includegraphics[width=.825\linewidth]{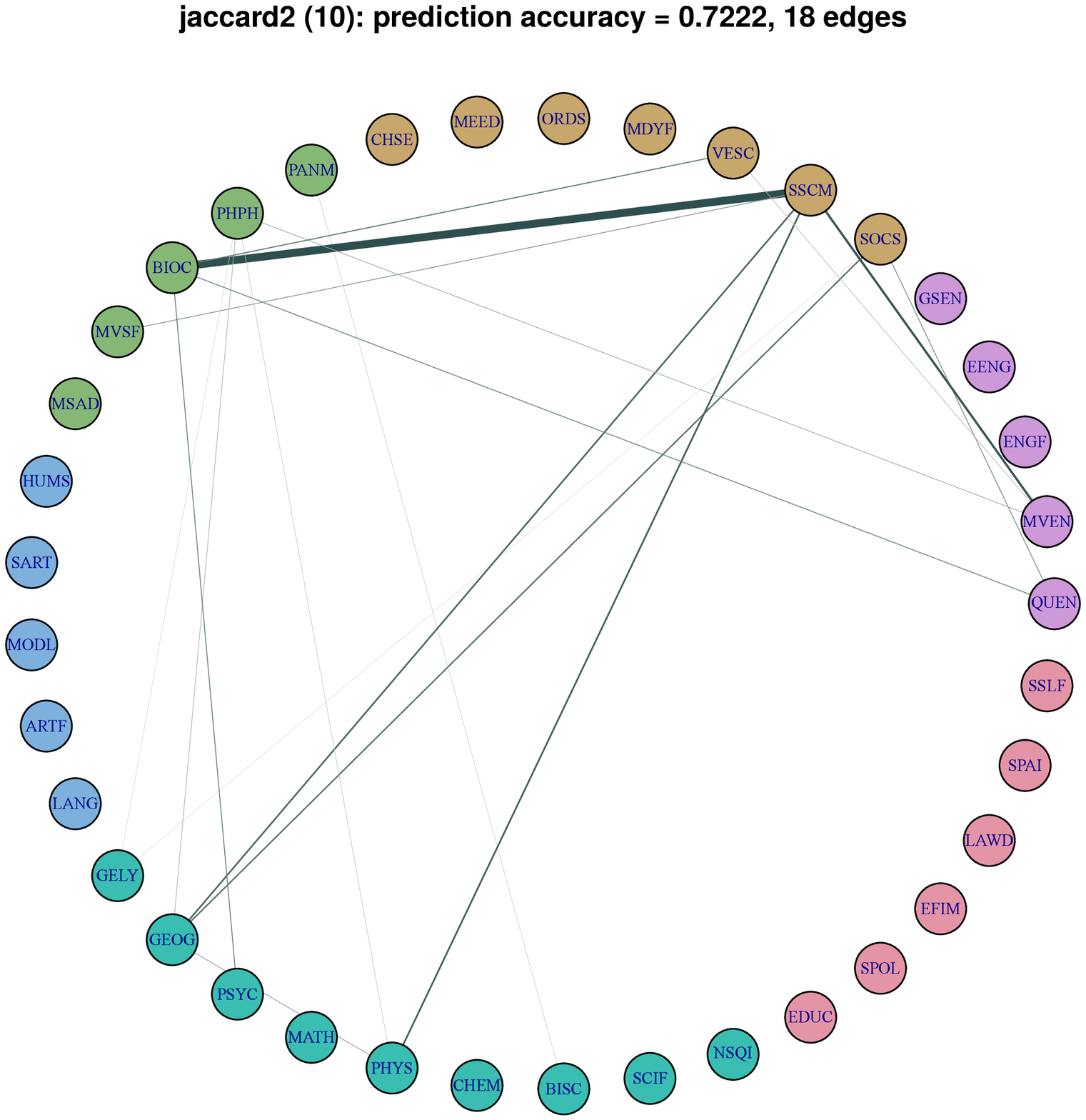} 
    \vspace{4ex}
  \end{minipage} 
  \begin{minipage}[b]{0.5\linewidth}
    \centering
    \includegraphics[width=.825\linewidth]{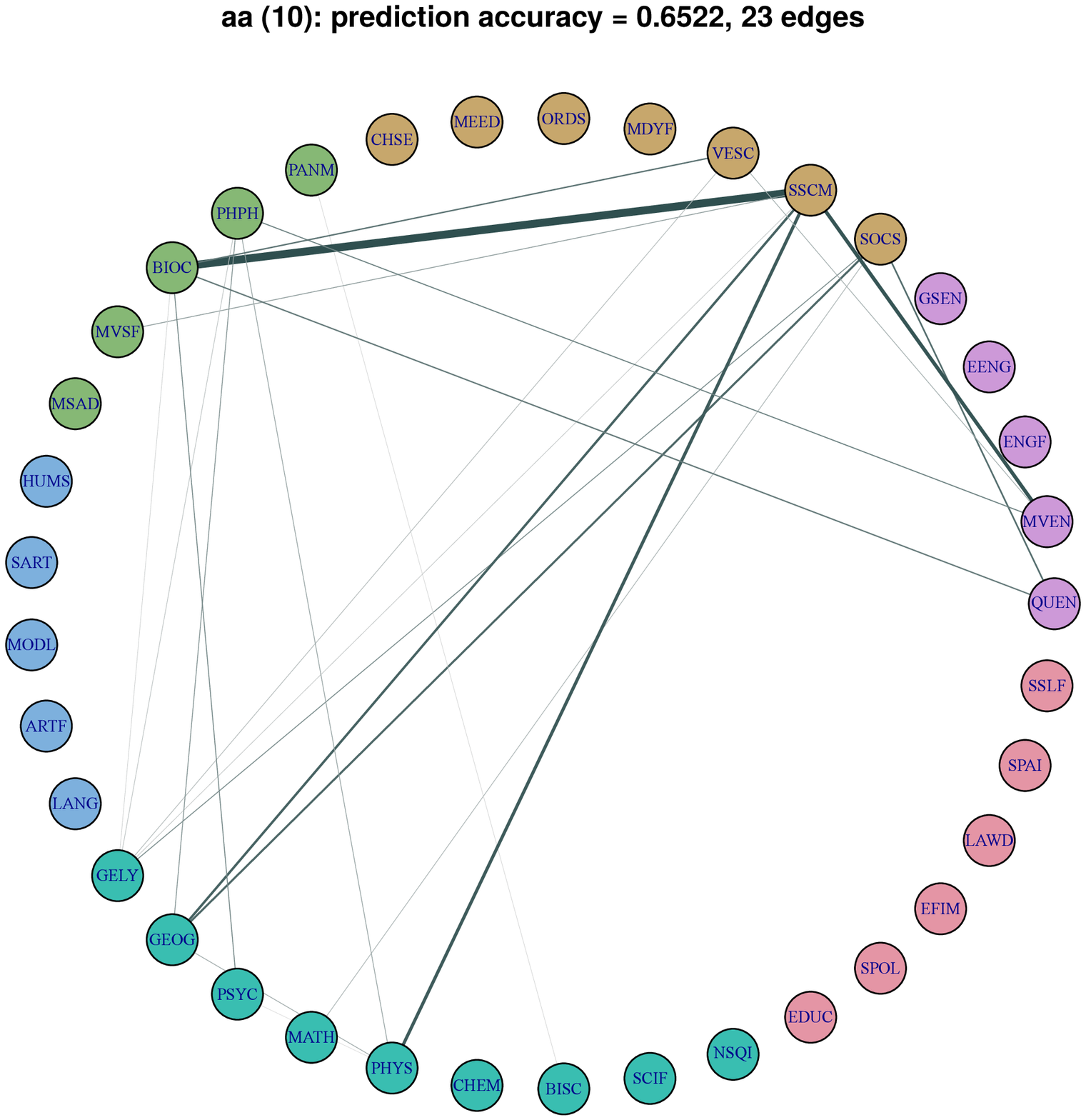} 
    \vspace{4ex}
  \end{minipage}
    \begin{minipage}[b]{0.5\linewidth}
    \centering
    \includegraphics[width=.825\linewidth]{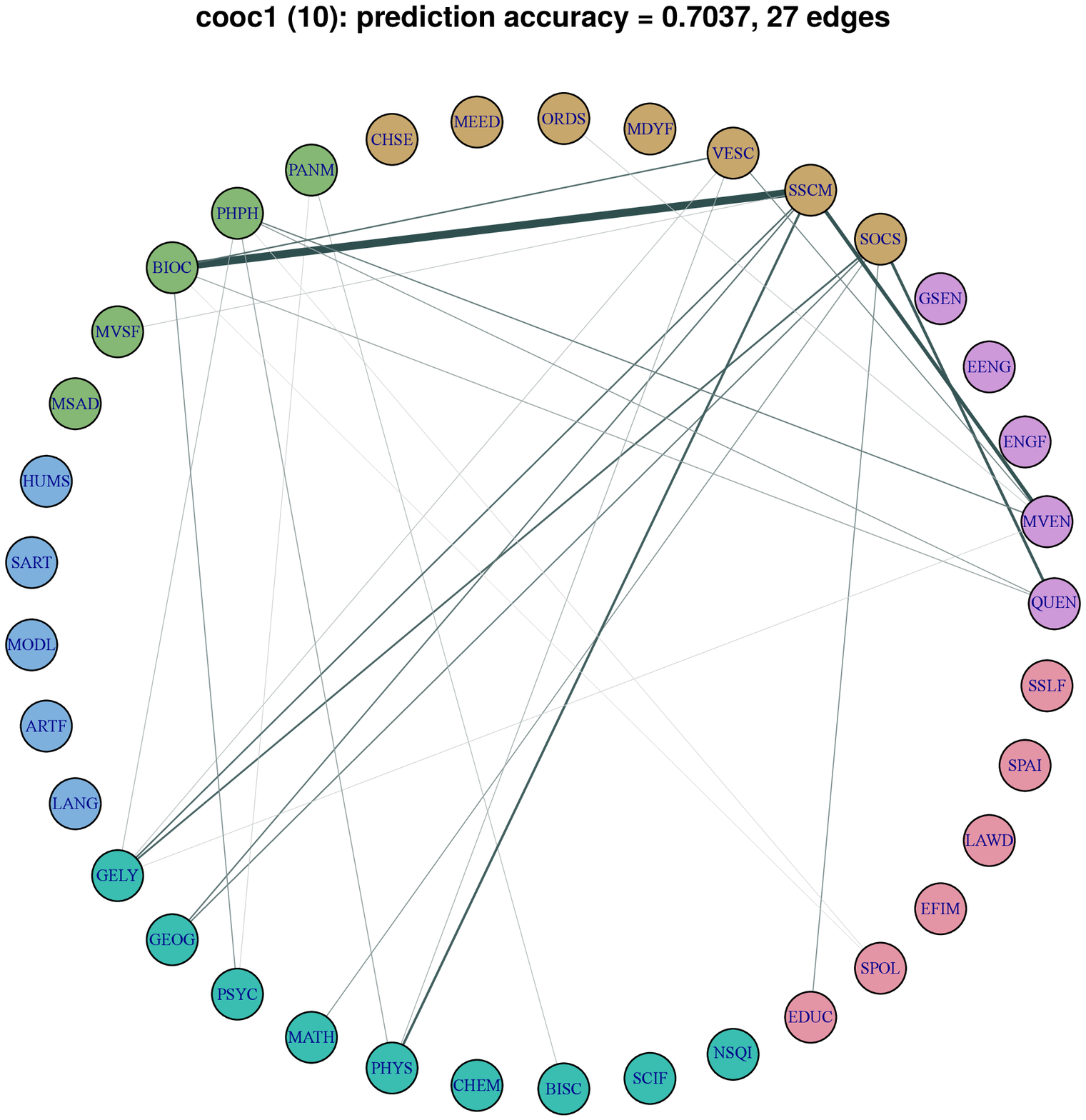} 
    \vspace{4ex}
  \end{minipage} 
    \begin{minipage}[b]{0.5\linewidth}
    \centering
    \includegraphics[width=.825\linewidth]{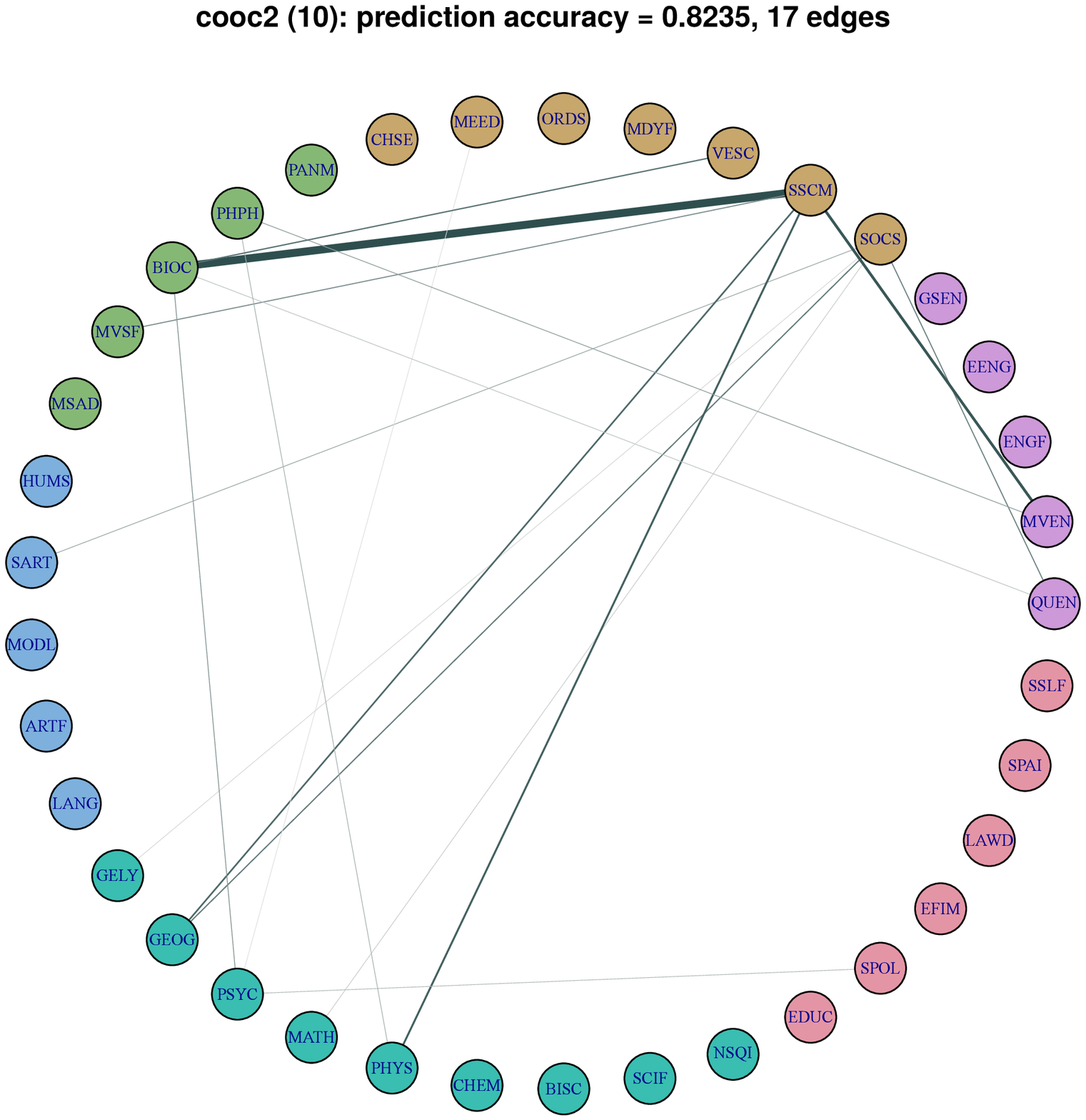} 
    \vspace{4ex}
  \end{minipage} 
  \begin{minipage}[b]{0.5\linewidth}
    \centering
    \includegraphics[width=.825\linewidth]{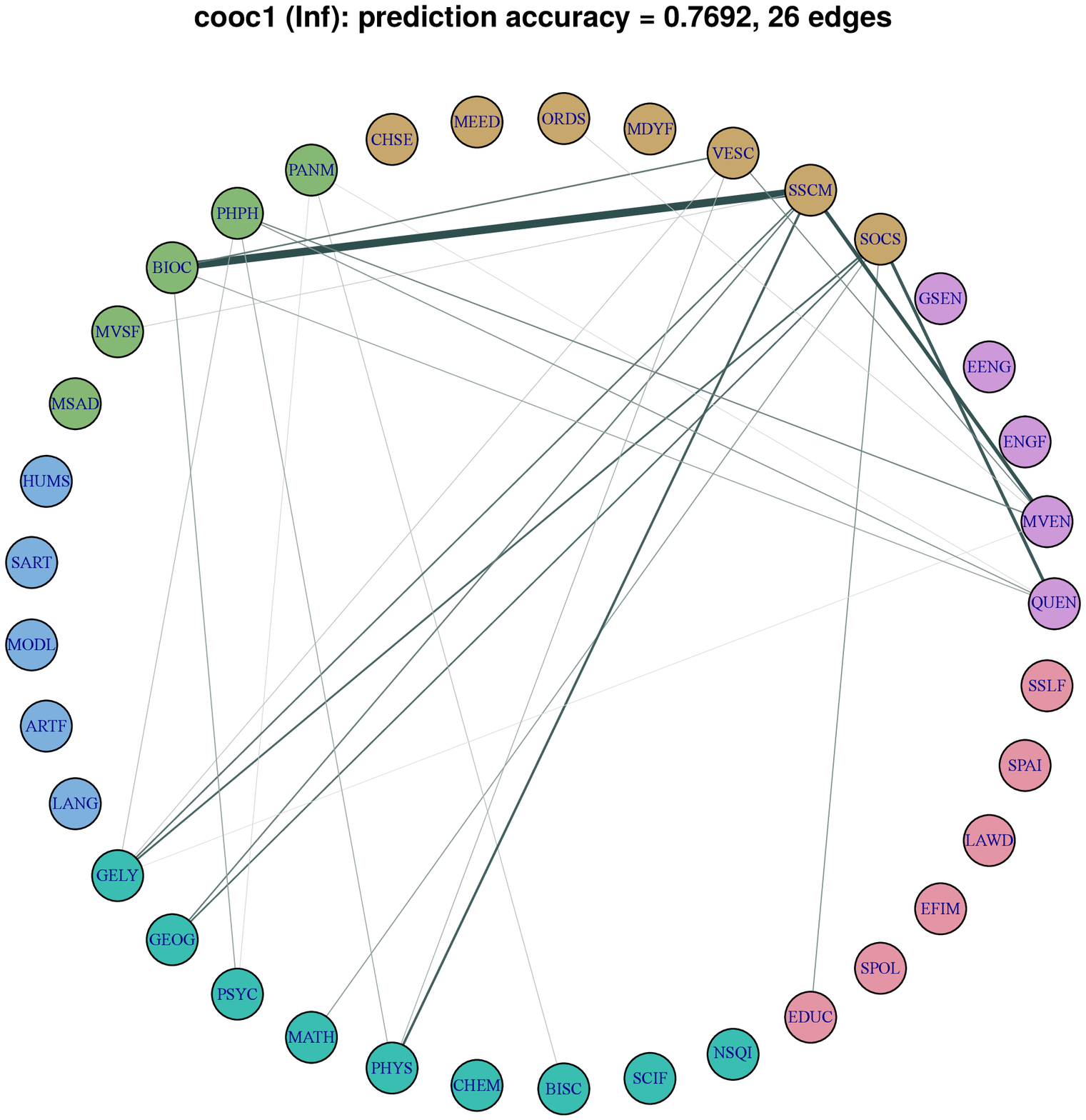} 
    \vspace{4ex}
  \end{minipage} 
\caption{Edges predicted indicating possible collaboration among School-level organisations using various similarity scores and $d$ (in parentheses) described in Section \ref{sec-similarity}. See Figure \ref{fig:nei:pred} for details about each graph.}
\label{fig:journal:pred}  
\end{figure}

In Figure~\ref{fig:nei:pred}, we present the edges predicted
with the similarity scores based on the co-authorship network with $p = 0.4$,
and in Figure~\ref{fig:journal:pred} those predicted with the similarity scores based on the bipartite network and $d = 10$,
in addition to the one returned with $\sigma^1_{\cooc}$ and $d = \infty$.  
Different node colours represent different Faculties to which Schools belong, 
and edge width is proportional to the edge weights $w_{kl}$ obtained in Step 2 of the proposed algorithm.  

Table~\ref{tab:nei-pred} shows that
the performance of the link prediction algorithm,
combined with the similarity scores based on the co-authorship network,
is not sensitive to the choice of the weights (a)--(d) nor the threshold ($p$):
all 16 combinations outperform the random choice, and do not differ too much among themselves.   
Only counting the length-2 geodesic path pairs, the score (a) predicts the most edges among them, 
and when no thresholding is applied ($p = 1$), all (a)--(d) select the same cohort of edges.  
From Figure \ref{fig:nei:pred}, it is observable that 
the four similarity scores still differ by preferring different edges. For instance,  
with (b) and (c), the edge between SSCM and GEOG is assigned a relatively larger weight than when (a) is used. 

It is evident that by taking into account the additional layer of information on journals enhances the prediction accuracy considerably, 
returning a larger proportion of true positives among a fewer number of predicted edges in general (thus fewer false positives).
In particular, combining the similarity measure $\sigma^1_{\cooc}$,
which takes into account the similarity among the journals as well,
with the choice $d \in \{\infty, 10\}$ returns a set of predicted edges
that is comparable to the set of edges predicted with the scores from Section \ref{sec-nei} in terms of its size,
while achieving higher prediction accuracy and recall.
Among possible values for $d$, most scores perform the best with $d =10$,
which aggregates the similarities between two individuals in forming School network edge weights,
provided that their geodesic distance in the co-authorship network is less than 10;
an exception is $\sigma^1_{\cooc}$, where slight improvement is observed with $d = \infty$.

For comparison, Table~\ref{tab:nei-pred} also reports the results 
from applying a modularity-maximising hierarchical community detection method 
to the School network constructed from $A^{\mathrm{train}}$.
Here, we assign an edge between Schools $k$ and $l$, $k, l \in \mathcal{O}$ 
with the number of publications between the researchers from the two Schools as its weight,
and the prediction is made by linking all the members (Schools) in the same communities.  
Modularity optimision algorithms are known to suffer from the resolution limit,
and strong connections among a small number of nodes in large networks are not well detected by such methods
\citep{Fortunato2007, AlzahraniHoradam2016}. 
Noting the nature of interdisciplinary research collaboration, 
which is often driven by a small number of individuals,
we choose to apply the community detection method to the School network of smaller size
rather than to the co-authorship network, 
following the approach described in (ii) at the beginning of Section \ref{sec:method}.

The optimal cut results in 21 different communities at the School level, which leads to too few predicted edges.
We therefore trace back in the dendrogram and show the results corresponding to the cases in which there are 5--8 communities.  
It is clearly seen from the outcome that our proposed method outperforms the community detection method 
regardless of the choice of similarity scores or other parameters. 
In fact, community detection often performs worse than random guessing in link prediction.
This may be attributed to modularity maximisation assuming
all communities in a network to be statistically similar \citep{newman2016}
whereas the PURE data set is highly unbalanced with regards to both the numbers of academic staff and publications 
at different Schools, see Figure \ref{fig:unbalance}.
On the other hand, our proposed method observes the potential for collaborative research at the individual level
and then aggregates the resulting scores to infer the interdisciplinary collaboration potential,
and hence can predict the links between e.g., a relatively small organisation (BIOC) and a large one (SSCM) as well as 
that between BIOC and another organisation of similar size (PSYC), see the bottom right panel of Figure \ref{fig:journal:pred}.

Our proposed method predicts edges which do not appear in the test data set. 
On one hand, this can be interpreted as false positive prediction
but on the other, it may be due to the time scale limitation, i.e.,  
these edges may appear after Year 2013, or the Schools connected still have the potential to form collaborative links which are yet to be realised.

\begin{figure}[htbp] 
	\begin{minipage}[b]{0.5\linewidth}
    \centering
    \includegraphics[width=.875\linewidth]{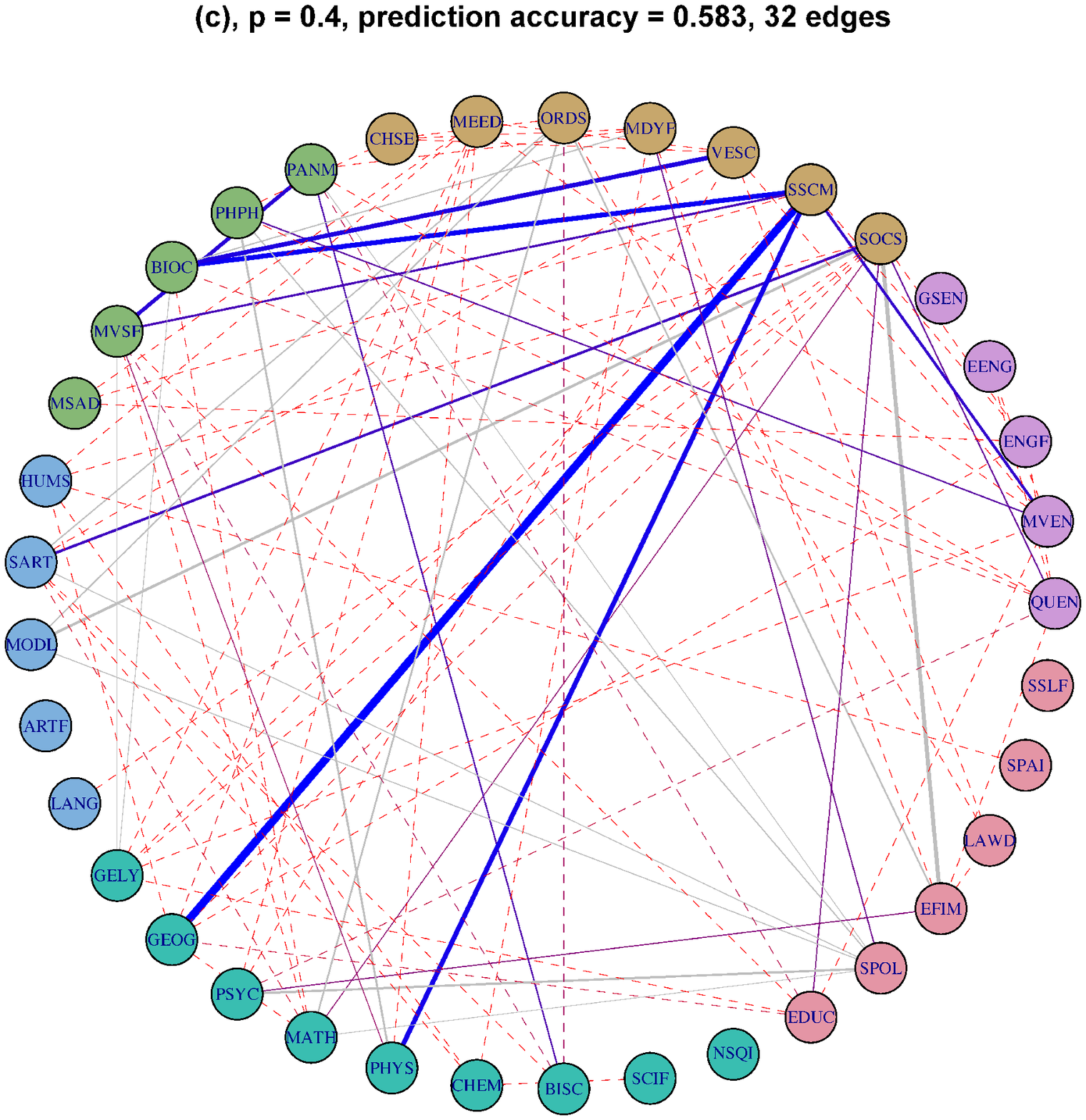} 
    \vspace{4ex}
  \end{minipage}
  \begin{minipage}[b]{0.5\linewidth}
    \centering
    \includegraphics[width=.825\linewidth]{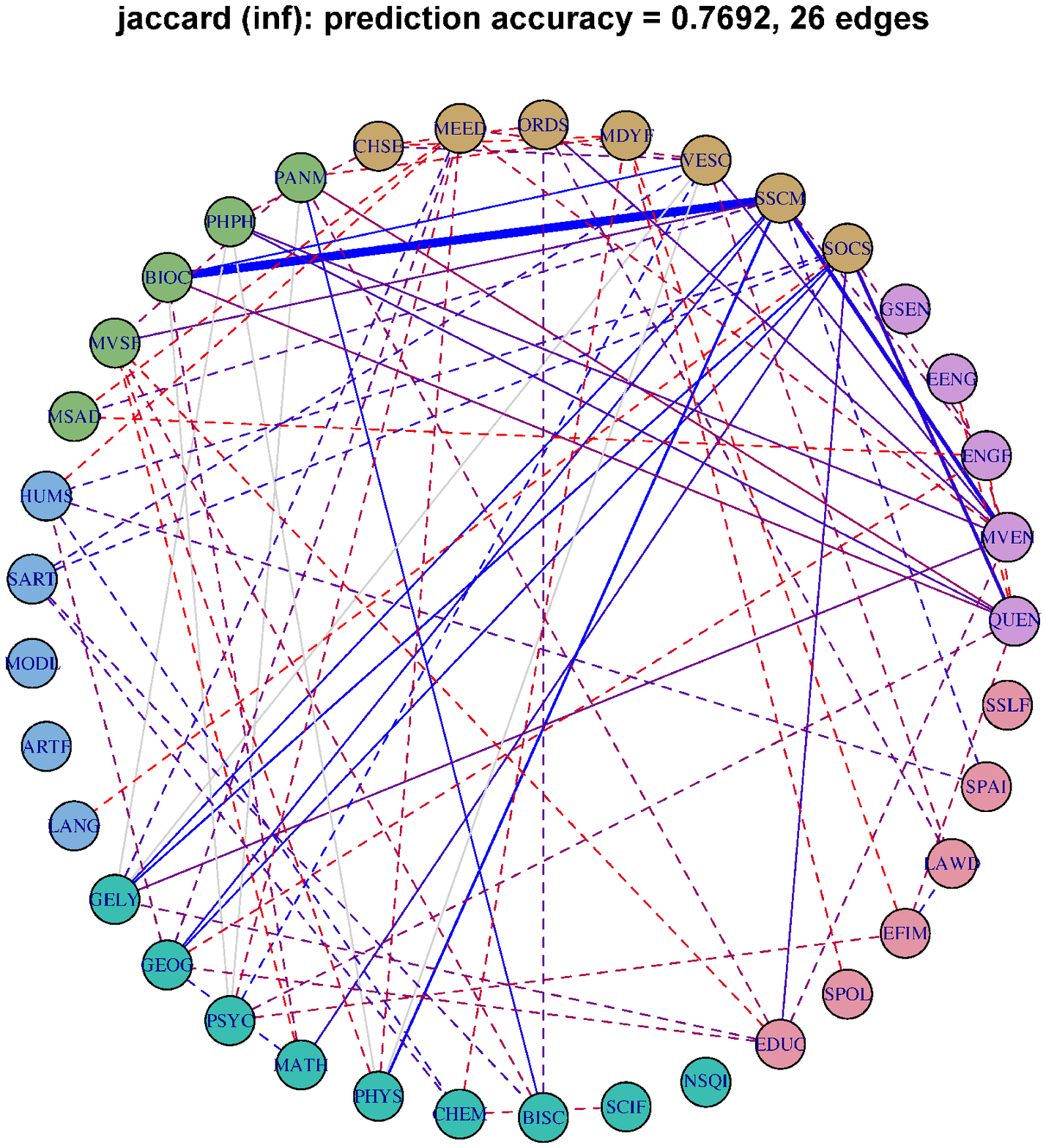} 
    \vspace{4ex}
  \end{minipage} 
 \caption{Edges in $E^{\mathrm{pred}} \cup E^{\mathrm{new}}$. Blue and red edges are in $E^{\mathrm{new}}$, and the bluer an edge is, the larger the corresponding weight that is computed using the test set; the redder an edge is, its test set weight is smaller. The edges in $E^{\mathrm{pred}} \setminus E^{\mathrm{new}}$ are in grey. The edges in $E^{\mathrm{pred}}$ are solid lines and their widths are proportional to $w_{kl}$, and the ones in $E^{\mathrm{new}} \setminus E^{\mathrm{pred}}$ are dashed lines. 
The left panel is based on the similarity score (c) with $p = 0.4$ described in Section \ref{sec-nei}, 
and the right panel is based on $\sigma^1_{\cooc}$ with $d = \infty$ as described in Section \ref{sec-similarity}.}
\label{fig:full.compare}  
\end{figure}

Figure~\ref{fig:full.compare} shows both the predicted edges (solid) and 
those which are in $E^{\mathrm{new}}$ but not among the predicted ones (false negatives, dashed). 
Edge width is proportional to the corresponding weight for $(k, l) \in E^{\mathrm{pred}}$. 
For the false negative edges, we assign a very small value (0.2) as their edge weights and add 0.2 to all other edge weights to make the visualisation possible.  
In addition, we use weights computed in the same manner but with the test data to colour the edges: the bluer an edge is, 
the greater the association is between the pairs of Schools connected in the test set, 
while the red edges indicate weaker association; grey ones are falsely predicted ones ($E^{\mathrm{pred}} \setminus E^{\mathrm{new}}$).  
In the figure, many of the predicted edges are more towards blue on the colour spectrum, 
while the majority of missing edges are in red, implying that the methodology is able to identify the pairs of Schools 
that develop significant collaboration in the test period.  

\section{Discussion}
\label{sec:discussion}

In this paper, we tackle the problem of predicting potential interdisciplinary research 
by transforming it to a membership network link prediction problem. 
Two types of similarity scores have been proposed in this paper, one employing 
only the co-authorship network and the other integrating additional information 
which is naturally available for the research output data. 
As expected, when we have more information in hand, the prediction accuracy improves. 
Within each type of scores, different choices of scores or parameters
do not differ by much in their performance when applied to the PURE data set.
However, this does not guarantee that the same robustness can be expected when different data sets are used. 

We would like to suggest that the practitioners make their own choice according to the aim of the analysis, 
and different behaviours of different metrics used may reflect the underlying properties of specific data set. 
For example, when using the co-author relationship only, 
if we also care about the amount of joint publications, 
then the similarity score (b) is more suitable. 
When additional information is available, $\sigma^1_{\cooc}$ 
returns the best prediction accuracy by taking into account 
not only those journals directly shared by two individuals, 
but also the journals which are similar to them.
Also, the scores proposed in Section \ref{sec-similarity} 
tend to return fewer edges and, consequently, 
fewer false positives which, for some applications, 
may be a more important criterion than the measure of prediction accuracy used in this paper.

We would also like to point out one main limitation of this paper.  
The problem here is to predict linkage between disciplines within a university.
However, due to the lack of information, 
it is not possible to map all individuals to disciplines
and therefore we equate disciplines with academic organisations within the university.
In most situations, this remedy works well, especially in traditional disciplines such as 
civil engineering, pure mathematics and languages, among others, 
which are all categorised well within the School framework.  
Relatively newer disciplines, however, do not have clear School boundaries, 
e.g., there are statisticians working in the School of Mathematics, 
School of Social and Community Medicine and School of Engineering. 
This situation on the other hand, also means mathematics, public health and engineering 
have shared interests in the modern world.  

Finally, the paper focuses on predicting academic collaboration links from the co-authorship network
but we would like to point out that the proposed method and similarity scores \emph{per se} 
are not limited to a single organisation or, indeed, an application area.  
For example, we may suggest interaction between different communities 
based on their members' Facebook networks, 
using both Facebook friend lists and additional information such as their taste in music or films.

\section*{Acknowledgements}

We thank the PURE team and the Jean Golding Institute at the University of Bristol for providing the data set.  
We thank Professor Jonathan C. Rougier for all the constructive discussions, comments and his input in the data analysis.  
We also thank the Editor and the two referees for their constructive suggestions.

\appendix
\section*{Appendix}

We provide in Table \ref{tab:org.key} the full names of the academic organisations
at the University of Bristol, supplementing Table \ref{tab:hier}.

\begin{table*}[htbp]
\begin{center}
\begin{tabular}{lp{5in}}
\underline{\underline{UNIV}} & \underline{\underline{University of Bristol}}\\
\underline{FENG} & \underline{Faculty of Engineering} \\
MVEN & Merchant Venturers' School of Engineering \\
& (changed to School of Computer Science, Electrical and \\
& Electronic Engineering, and Engineering Mathematics) \\
QUEN & Queen's School of Engineering \\
& (changed to School of Civil, Aerospace and Mechanical Engineering) \\
EENG & Department of Electrical \& Electronic Engineering \\
GSEN & Graduate School of Engineering \\
ENGF & Engineering Faculty Office \\
\underline{FMDY} & \underline{Faculty of Health Sciences} \\
ORDS & Oral \& Dental Sciences \\
SOCS & Clinical Sciences (changed to Population Health Sciences) \\
SSCM & Social and Community Medicine (changed to Translational Health Sciences) \\
VESC & Veterinary Sciences \\
MDYF &  Health Sciences Faculty Office\\
MEED & Centre for Medical Education\\
CHSE & Centre for Health Sciences Education \\
\underline{FMVS} & \underline{Faculty of Biomedical Sciences} \\
BIOC & Biochemistry \\
PANM & Cellular and Molecular Medicine \\
PHPH & Physiology, Pharmacology \& Neuroscience \\
MVSF & Biomedical Sciences Faculty Office \\
MSAD & Biomedical Sciences Building \\
\underline{FOAT} & \underline{Faculty of Arts} \\
HUMS & Humanities \\
MODL & Modern Languages \\
SART & Arts \\
ARTF & Faculty Office Arts Faculty Office \\
LANG & Centre for English Language and Foundation Studies \\
\underline{FSCI} & \underline{Faculty of Science} \\
BISC & Biological Sciences \\
CHEM & Chemistry \\
GELY & Earth Sciences \\
GEOG & Geographical Sciences \\
MATH & Mathematics \\
PHYS & Physics \\
PSYC & Experimental Psychology \\
NSQI & Centre for Nanoscience and Quantum Information \\
SCIF & Science Faculty Office \\
\underline{FSSL} & \underline{Faculty of Social Sciences and Law} \\
EDUC & Graduate School of Education \\
EFIM & Economics, Finance and Management \\
LAWD & University of Bristol Law School \\
SPAI & Sociology, Politics and International Studies \\
SPOL & Policy Studies \\
SSLF & Social Sciences and Law Faculty Office \\
\end{tabular}
\caption{Abbreviations and full names of the academic organisations \label{tab:org.key}}
\end{center}
\end{table*}

\end{document}